\documentclass[journal]{IEEEtran}
\ifCLASSINFOpdf
\else
\fi

\makeatletter

\newcommand{\Rmnum}[1]{\expandafter\@slowromancap\romannumeral #1@}
\makeatother

\usepackage{graphicx}
\usepackage{amsmath}
\usepackage{CJK}

\usepackage{multirow}
\usepackage{booktabs}
\usepackage{algorithm}
\usepackage{subfigure}
\usepackage{epstopdf}
\usepackage{graphics}
\usepackage{epsfig}
\usepackage{psfrag}
\usepackage{overpic}
\usepackage{amsfonts}
\usepackage{stfloats}
\usepackage{float}   
\usepackage{algpseudocode}
\usepackage{indentfirst}
\usepackage{array}
\usepackage{url}
\usepackage{cite}

\begin{document}

\title{Improving Power System State Estimation Based on Matrix-Level Cleaning}

\author{Haosen Yang\textsuperscript{1},
        Robert C.~Qiu\textsuperscript{1},~\IEEEmembership{IEEE~Fellow}, Lei Chu\textsuperscript{1}~\IEEEmembership{IEEE~Student~Member}, Tiebin Mi\textsuperscript{1}~\IEEEmembership{IEEE~Member}, Xin Shi\textsuperscript{1}~\IEEEmembership{IEEE~Student~Member}, Chaoyuan Mary Liu\textsuperscript{2} 
\thanks{\textsuperscript{1} Department of Electrical Engineering, Center for Big Data and Artificial Intelligence, Shanghai Jiaotong University, Shanghai 200240, China.}
\thanks{\textsuperscript{2} Department of Mathematics and Statistics, Eastern Kentucky University, Richmond, KY 40475, USA.}
\thanks{Email: Robert C. Qiu: rcqiu@sjtu.edu.cn; Haosen Yang: 31910019@sjtu.edu.cn.}
}

\maketitle

\begin{abstract}
  Power system state estimation is heavily subjected to measurement error, which comes from the noise of measuring instruments, communication noise, and some unclear randomness. Traditional weighted least square (WLS), as the most universal state estimation method, attempts to minimize the residual between measurements and the estimation of measured variables, but it is unable to handle the measurement error. To solve this problem, based on random matrix theory, this paper proposes a data-driven approach to clean measurement error in matrix-level. Our method significantly reduces the negative effect of measurement error, and conducts a two-stage state estimation scheme combined with WLS. In this method, a Hermitian matrix is constructed to establish an invertible relationship between the eigenvalues of measurements and their covariance matrix. Random matrix tools, combined with an optimization scheme, are used to clean measurement error by shrinking the eigenvalues of the covariance matrix. With great robustness and generality, our approach is particularly suitable for large interconnected power grids. Our method has been numerically evaluated using  different testing systems, multiple models of measured noise and matrix size ratios.
\end{abstract}


\IEEEpeerreviewmaketitle

\section{Introduction}
\IEEEPARstart{P}{ower} system state estimation aims to estimate state variables from measurement data corrupted with noise, and it plays an important role in power system operations, such as optimal power flow, stability analysis, and economic dispatch. With the development of energy management systems (EMS) and smart grids, the requirement for accurate operating parameters has been increasing greatly \cite{new_challenge}.
\\\indent Conventional state estimation is mainly based on WLS, which solves the normal equation iteratively by Gauss-Newton method \cite{proceedings}. Despite long and wide applications of WLS, there is increasing concern for its accuracy and robustness. The objective function of WLS is the residual between measurements and the estimation of measured variables. Minimizing the residual brings the estimation of measured variables close to measurements. However, approaching measurements is dissimilar from approaching true values, since measurements are corrupted by the noise of measuring instruments, communication noise and random fluctuations. So only minimizing the residual but disregarding measurement error results in a certain degree of estimated error. Besides, WLS heavily suffers from the ill-conditional gain matrix and bad data, and it is sensitive to the initialization of state variables.
\\\indent In recent years, many researchers have been searching novel approaches to improve power system state estimation \cite{bilinear,mimutes, huber,LAV,SSR,autoencoder,combined,factor_WLS}. In \cite{LAV}, a least-absolute-value (LAV) guided estimator was proposed, which is more robust and exhibits many advantages for phasor measurements. In \cite{SSR}, an iterative $l_{1}$-$l_{2}$ mixed convex programming was used for state estimation by linearizing the nonlinear physical equations. In \cite{autoencoder}, an autoencoder based pre-filtering was proposed to clean measurement noise and remove gross errors. But as a deep learning method, autoencoder spends too much time in off-line training. In \cite{combined} and \cite{factor_WLS}, a two-stage state estimation method was designed, in which the measured variables are transformed into a new group of variables at first so that the measurement model in the second stage is linear.
\\\indent By reviewing, even though many studies investigated new approaches to improve state estimation, few researchers considered directly processing measurement error in any systematic way. Our work aims to fill this gap and improve the accuracy of state estimation. Because of the strong randomness and  unclear influence of multiple noise, measurement error is difficult to be handled in vector form or single-value form in the past. Nevertheless, along with the well-established research line of random matrix theory (RMT), some deterministic properties are workable when gathering the fully stochastic measurement error in matrix form. For instance, the M-P law, proposed in \cite{mp}, reveals that the eigenvalues of a Gaussian covariance matrix asymptotically converge to a deterministic probability distribution. Inspired by this, based on RMT and an optimization scheme, this paper proposes a two-stage state estimation method, in which we process measurements by RMT firstly and then use WLS to estimate state variables. RMT, aiming to extract insightful information from eigenvalues distributions of large covariance matrices, has emerged as a particularly useful framework for many theoretical questions associated with high-dimensional big data analytics \cite{qiu-book1}\cite{qiu-paper2}. It has been successfully applied in quantum physics \cite{quantum}, wireless communication \cite{chu_eigen}, and signal processing \cite{RMT-wai1}, for its remarkable effect in dealing with measurement noise in matrix-level. And large amounts of measurement data collected from monitoring systems provide a new opportunity for proper applications of RMT in power systems, including event detection \cite{shixin} and correlation analysis \cite{RMT-correlation}. In this work, an optimization framework, derived from RMT, is used to clean measurement error by filtering the eigenvalues of the covariance matrix, after forming a Hermitian matrix which is used to establish an invertible relationship between the eigenvalues of measurements and those of their covariance matrix. To my best knowledge, it is the first time for RMT to be applied in power system state estimation.
\\\indent The contributions of this paper are listed as follows:
\\\indent (1) A crucial drawback of traditional state estimation methods is clearly analyzed that WLS does not take the effect of measurement error into account. And the association among measurement error, residual and estimated error is quantitatively discussed.
\\\indent (2) We propose a Hermitian matrix construction method to extend the application scope of previous noise-cleaning methods. Most of previous papers (e.g., \cite{cleaning,previous,LW4}), aimed to clean measurement noise based on RMT, only involved eliminating noise of covariance matrices, but they did not intend to clean errors in original data of power systems. So the Hermitian matrix construction is designed to enable the RMT based error-cleaning scheme to adapt to measurements of power systems.
\\\indent (3) A two-stage state estimation framework is proposed, in which a matrix-level cleaning method is used at first to obtain more reasonable measured values, and then WLS is employed to eventually calculate the state vector. This framework solves what the previous state estimation methods neglect, and greatly improves the accuracy of state estimation.
\\\indent The rest of this paper is organized as follows, section \Rmnum{2} introduces the problem statement and current drawbacks. Section \Rmnum{3} talks about the proposed methodology. Section \Rmnum{4} is case studies, and section \Rmnum{5} summarizes our work.
\section{Problem Statement}
\subsection{State Estimation}
The measurement model of a power system is:
\begin{equation}\label{1}
\bf{z=h(x)+e}
\end{equation} where $\bf{z}$ denotes the measurement vector. When $\bf{z}$ comes from SCADA, it usually contains power flows, nodal injective power and nodal voltage magnitudes. $\bf{x}$ represents the state vector including nodal voltage magnitudes and voltage angles. $\bf{h}(\cdot)$ denotes the nonlinear function relating $\bf{z}$ to $\bf{x}$. And $\bf{e}$ is the vector of measurement error whose elements are generally assumed to follow Gaussian distributions. State estimation is usually considered as a typical WLS problem, in which the objective function is:
\begin{equation}\label{2}
J=||{\bf{(z-h(x))}}^{T}{\bf{W(z-h(x))}}||^{2}
\end{equation} where $||\cdot||^{2}$ is the $l_{2}$ norm. ${\bf{W}}=diag\{\sigma_{1}^{2},1/\sigma_{2}^{2},1/\sigma_{3}^{2},\cdot \cdot \cdot1/\sigma_{n}^{2}\}$ is a diagonal weighted matrix whose elements are reciprocals of the variance of measurement errors. $\sigma_{i}^{2}$ denotes the variance of the Gaussian error for the $i$-$th$ measured variable.
\\\indent Function (\ref{2}) can be minimized by iteratively solving the well-known normal equation:
\begin{equation}\label{3}
\bf{H^{T}WH\Delta x=HW(z-h(x))}
\end{equation} where $\bf{H=\partial h(x)/\partial x}$ is the jacobian matrix of $\bf{h(x)}$ w.r.t $\bf{x}$. Upon convergence, an estimated state vector $\bf{\hat{x}}$ is obtained, and some techniques on gross error detection will be operated.
\subsection{Issue of Residual}
It is clear that we have three vectors of measured variables: measurements $\bf{z}$, true values $\bf{h(x)}$ and estimated values $\bf{\hat{z}=h(\hat{x})}$. The residual is defined as the difference vector between measurements $\bf{z}$ and estimated values $\bf{h(\hat{x})}$:
\begin{equation}\label{4}
\bf{r=z-h(\hat{x})}
\end{equation} The vector of measurement error is the difference between measurements $\bf{z}$ and real values $\bf{h(x)}$:
\begin{equation}\label{5}
\bf{e=z-h(x)}
\end{equation}
The vector of estimated error is the difference between real values $\bf{h(x)}$ and estimated values $\bf{h(\hat{x})}$:
\begin{equation}\label{add6}
\bf{R_{e}=h(\hat{x})-h(x)}
\end{equation} To guide the optimization process, a distance function $L(\cdot)$ is required to measure the scale of these distance vectors, such as $l_{2}$ norm, $l_{1}$ norm, etc. The distance function $L(\cdot)$ must satisfy three requirements: (1) Triangular inequality, i.e., $L(a+b)\leq L(a)+L(b)$; (2) Symmetry, $L(-a)=L(a)$; (3) $L(\cdot)\geq0$. So function (\ref{add6}) can be resolved:
\begin{equation}\label{6}
\begin{split}
L({\bf{WR_{e})}}&=L({\bf{W(h(\hat{x})-h(x)))}} \\
&=L({\bf{W(h(\hat{x})-z+z-h(x)))}}   \\
& \leq L({\bf{W(z-h(\hat{x}))}}+L({\bf{W(z-h(x)))}} \\
& = L({\bf{W(r+e))}}
\end{split}
\end{equation} where $\bf{W}$ is defined similarly as that in function (\ref{2}). According to function (\ref{6}), it is obvious that the estimated error $\bf{R_{e}}$ is mainly influenced by two crucial components: the residual $\bf{r}$ and the measurement error $\bf{e}$. However, WLS fails to consider the influence of $\bf{e}$, thus the estimation effect is significantly limited by its nature.
\begin{figure}
  \centering
  \includegraphics[height=2.3in,width=3.3in]{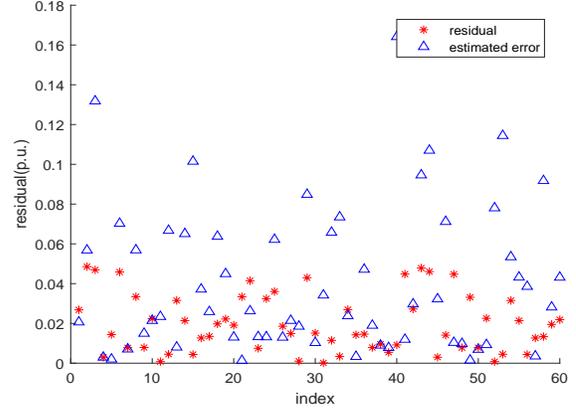}\\
  \caption{The comparison between the residual and the estimated error of every variable. The index 1 to 30 in the abscissa represents the magnitude of every nodal voltage, while 31 to 60 means nodal voltage angles. This case is tested in IEEE 30-bus system, and all the variables are standardized.}\label{Fig.01}
\end{figure}
\\\indent To support the above analysis, a quick test using WLS is operated in IEEE 30-bus system, which contains 254 measured variables (including power flows, nodal injective power, and voltage magnitudes) and 60 state variables (nodal voltage magnitudes and angles) in total. In the simulation, Gaussian measurement errors with zero mean are added, whose variance is 5\% of original power flows, as well as 1\% for original voltage magnitudes. The initial values of voltage magnitudes and angles are randomly sampled from the Gaussian distribution $\mathcal{N}(1, 0.05)$ and $\mathcal{N}(0, 0.157)$.
\\\indent The residual and the estimated error of every state variable are plotted in Fig. \ref{Fig.01}. Though the residuals of most of variables have been shrunk well, the estimation of partial measured variables is still unsatisfactory, showing that it is extremely unfeasible to ignore the existence of measurement error. To address this issue, we propose a data-driven method to process measurement error.
\section{Proposed Methodology}
Our method to clean measurements in matrix-level involves two main parts: Hermitian matrix construction and RMT based error cleaning (RBEC). The Hermitian matrix construction is responsible for recovering the eigenvalues of the measurement matrix from those of its covariance matrix. And the purpose of RBEC is to clean the covariance matrix through shrinking its eigenvalues.
\\\indent \textit{Assumptions}: (1) The topology and parameters of the estimated power system are available, which is a necessary condition for estimating state variables.
\\\indent (2) The variance of measurement error can be estimated. This assumption, as same as most of studies about power system state estimation, arises from the fact that there are a large number of approaches to estimate the variance of measurement noise by pseudo-measurements (historical data), including direct method \cite{measurement}, statistical inference \cite{var-sta}, empirical Bayes estimation \cite{var-decon} and covariance matrix analysis \cite{var-covar1}\cite{var-covar2}, etc. For instance, the variance of the individual model $z_{i}=h(x_{i})+e_{i}$ can be written as $\sigma^{2}_{z}=\sigma^{2}_{h} +\sigma^{2}_{e}$, where $\sigma_{h}$ is the variance of normal fluctuations in operating state, and $\sigma_{e}$ denotes the variance of measurement error. So that the commonest method is performing a large number of independent repeated samplings over a time period during which the operating state does not change much ($\sigma_{h}\approx 0$). As for the bias of measurement error, it can also be estimated easily by calculating the average of the non-absolute error (i.e., $b_{i}=\mathbb{E}(z_{i}-\hat{z}_{i})$ ). Readers can refer to \cite{measurement} for more information about methods to estimate the variance and bias of measurement error.
\\\indent (3) The measurement noise is Gaussian distributed. This assumption is based on the central limit theorem that the sum of multiple independent random variables tends to be a Gaussian variable even if the respective variables are not Gaussian distributed. Measurement error is caused by the accumulation of uncertainties of measuring instruments, communication noise and other unclear randomness, so it is generally modeled as a Gaussian distribution. If measurement error is strictly Gaussian distributed, our method can obtain the optimal result. However, in practice, measurement error may do not rigorously follow a Gaussian distribution, because some indescribable factors may slightly change it. This minor violation of the Gaussian hypothesis will reduce the effectiveness of our approach. To empirically illustrate the performance in the case of non-Gaussian distributions, many experiments using different distributions of measurement error are tested in case studies.
\subsection{Normalization and Matrix Formation}
Our method cleans measurement noise in matrix-level, for it is discovered that Gaussian measurement error in matrix form possesses some excellent and analytical properties in the distribution of its eigenvalues. Before forming a matrix, we normalize the measurement error of each measured variable by:
\begin{equation}\label{7}
\tilde{z_{i}}=\frac{z_{i}-b_{i}}{\sigma _{i}}=(z_{i}-b_{i})\sqrt{w_{i}}
\end{equation} where $z_{i}$ denotes the $i$-$th$ variable in the measurement vector, $\sigma_{i}$ represents the standard deviation of its measurement error $e_{i}$. $b_{i}$ is the bias of $e_{i}$, and $w_{i}=1/\sigma _{i}^{2}$ is the $i$-$th$ weighted coefficient in WLS. Then we utilize a split sample window to form an $N\times T (N<T)$ measurement matrix ${\bf{Z}} \in \mathbb{R}^{N\times T}$:
\begin{equation}\label{8}
\bf{Z}=[\tilde{z}_{k-N+1},\tilde{z}_{k-N+2}......\tilde{z}_{k-1},\tilde{z}_{k}]
\end{equation} where $\bf{\tilde{z}_{k}}$, by slight abuse of notations, is the normalized current measurement vector. $N$ represents the number of rows (the number of sample variables), specifically each row represents one variable sampled from SCADA. The number of columns $T$ denotes the number of continuous samplings over a period of time.
\\\indent Assuming the matrix of real values ${\bf{H}}\in \mathbb{R}^{N\times T}$ corresponding to $\bf{Z}$ is:
\begin{equation}\label{9}
\bf{H=[\widetilde{h(x)}_{k-N+1},\widetilde{h(x)}_{k-N+2}......\widetilde{h(x)}_{k-1},\widetilde{h(x)}_{k}]}
\end{equation} where $\bf{\widetilde{h(x)}_{k}}$ is the vector of real values after normalization. Then we have the model:
\begin{equation}\label{10}
\bf{Z=H+G}
\end{equation} where $\bf{G}$ is the normalized Gaussian matrix in which every entry follows an independent identical distribution (i.i.d) with zero mean and unit variance.
\subsection{Hermitian Matrix Construction}
The reason why we construct a Hermitian matrix is to utilize its property: The absolute values of the eigenvalues of a Hermitian matrix are equal to their corresponding singular values. Here the Hermitian matrix ${\bf{D_{Z}}} \in \mathbb{R}^{(N+T)\times (N+T)}$ is constructed by:
\begin{equation}\label{11}
\bf{D_{Z}}=\begin{bmatrix}
& \bf{Z} \\
\bf{Z^{H}} &
\end{bmatrix}
\end{equation} where ${\bf{Z}}^{H}$ is the associate matrix (conjugate transpose) of $\bf{Z}$. Then the covariance matrix of $\bf{D_{Z}}$ is:
\begin{equation}\label{12}
\bf{F_{Z}}=\begin{bmatrix}
\bf{ZZ^{H}} &  \\
 &  \bf{Z^{H}Z}
\end{bmatrix}
\end{equation} Besides, another important property of the matrix $\bf{D_{Z}}$ is that if $\lambda _{D}$ is an eigenvalue of $\bf{D_{Z}}$, its contrary value $-\lambda _{D}$ must be another eigenvalue. According to this and the property of Hermitian matrices introduced above, the relationship between the eigenvalues of $\bf{D_{Z}}$ and $\bf{F_{Z}}$ is:
\begin{equation}\label{13}
\lambda _{D_{Z}}=\pm \sqrt{\lambda _{F_{Z}}}
\end{equation} where $\lambda _{D_{Z}}$ and $\lambda _{F_{Z}}$ denote the eigenvalues of $\bf{D_{Z}}$ and $\bf{F_{Z}}$ separately. It is easy to find that ${\bf{ZZ}}^{H}$ and ${\bf{Z}}^{H}{\bf{Z}}$ have the same nonzero eigenvalues. And ${\bf{Z}}^{H}{\bf{Z}}$ has other $T-N$ zero eigenvalues when $N<T$, so that the eigenvalues of $\bf{F_{Z}}$ and ${\bf{ZZ}}^{H}$ are reversibly related:
\begin{equation}\label{13add}
\lambda_{F_{Z}}=\left\{\begin{matrix}
\lambda_{ZZ^{H}} & \lambda_{F_{Z}}\neq 0 \\
0 & \lambda_{F_{Z}}=0
\end{matrix}\right.
\end{equation} Same as $\bf{Z}$, for the matrix of real values $\bf{H}$, we have:
\begin{equation}\label{14}
\bf{D_{H}}=\begin{bmatrix}
 & \bf{H} \\
\bf{H^{H}} &
\end{bmatrix}, \quad
\bf{F_{H}}=\begin{bmatrix}
\bf{HH^{H}} &  \\
 &  \bf{H^{H}H}
\end{bmatrix}
\end{equation} and:
\begin{equation}\label{15}
\begin{split}
&\lambda _{D_{H}}=\pm \sqrt{\lambda _{F_{H}}} \\
\lambda_{F_{H}} &=\left\{\begin{matrix}
\lambda_{HH^{H}} & \lambda_{F_{H}}\neq 0 \\
0 & \lambda_{F_{H}}=0
\end{matrix}\right.
\end{split}
\end{equation} where $\lambda _{D_{H}}$, $\lambda _{F_{H}}$ and $\lambda_{HH^{H}}$ denote the eigenvalues of $\bf{D_{H}}$, $\bf{F_{H}}$ and ${\bf{HH}}^{H}$, respectively.
\begin{figure}
  \centering
  \includegraphics[height=1.4in,width=2.6in]{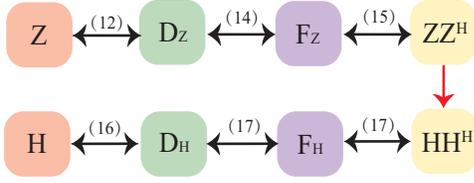}\\
  \caption{The flow of the invertible relationship between the eigenvalues of different matrices. The bracket above arrows is the function this step corresponds to. }\label{Fig.02}
\end{figure}
\\\indent The overall computing flow of eigenvalues is shown in Fig. \ref{Fig.02}, in which we aim to connect the eigenvalues of $\bf{Z}$ and $\bf{H}$. For now, except for the calculation from ${\bf{ZZ}}^{H}$ to ${\bf{HH}}^{H}$ (the red arrow in Fig. \ref{Fig.02}), all procedures in this eigenvalues flow are available.

\subsection{RMT Based Error Cleaning}
\begin{figure}
  \centering
  \includegraphics[height=3.1in,width=3.3in]{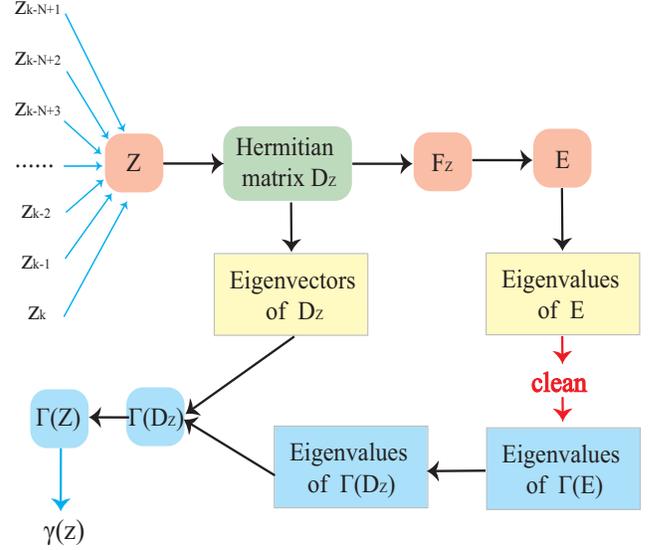}\\
  \caption{The entire process of RBEC.}\label{Fig.03}
\end{figure}

Now we introduce how to estimate the eigenvalues of ${\bf{HH}}^{H}$ from ${\bf{ZZ}}^{H}$ by RBEC and how to estimate the real values of measured variables. To facilitate reading, some related mathematical equations of RMT are presented in appendix.~A.
\\\indent Now we denote the covariance matrices by ${\bf{E=ZZ}}^{H}$ and ${\bf{C=HH}}^{H}$. Then model (\ref{10}) can be expressed in the form of covariance matrices:
\begin{equation}\label{bujiao}
\begin{split}
\bf{E} &={\bf{(H+G)(H+G)}}^{H}\\
  &={\bf{C+GG}}^{H}+{\bf{HG}}^{H}+{\bf{GH}}^{H} \\
  &=\bf{C+B}
\end{split}
\end{equation} where we denote the last three terms ${\bf{GG}}^{H}+{\bf{HG}}^{H}+ {\bf{GH}}^{H}$ by $\bf{B}$. Let $c_{i}$, $\bf{v^{}_{li}}$ and $\bf{v_{ri}}$ denote the $i$-$th$ eigenvalue, left and right eigenvector of $\bf{C}$, respectively. $\lambda_{i}$, $\bf{u^{}_{li}}$ and $\bf{u_{ri}}$ denote the $i$-$th$ eigenvalue, left and right eigenvector of $\bf{E}$, respectively. $\xi_{i}$ is the $i$-$th$ estimated eigenvalue corresponding to $\lambda_{i}$. The objective function of our RBEC is:
\begin{equation}\label{24}
\begin{split}
\underset{\Gamma(\bf{E})}{\min} \quad & ||\Gamma\bf{(E)-C}||^{2} \\
= \quad & Tr[(\Gamma\bf{(E)-C)}^{2}] \\
\end{split}
\end{equation} where $\Gamma(\bf{E})$ is the estimation of $\bf{C}$ from $\bf{E}$. Since $\Gamma(\bf{E})$ and $\bf{C}$ are both symmetric matrices, the $l_{2}$ norm of $\Gamma\bf{(E)-C}$ is equal to the trace of its square, as the second line in function (\ref{24}) shows. RMT shows the asymptotically deterministic property in the eigenvalues distribution of Gaussian error matrices by using multiple transforms (see appendix. A). So the eigenvalues distributions of matrices in $\bf{B}$ can be analytically connected with $\bf{E}$ and $\bf{C}$. Thus the trace (equal to the sum of its eigenvalues) of the matrix $\bf{C}$ can be approximated.
\\\indent To solve this optimization problem, similar to the family of rotational invariant methods \cite{rie}, we assume that $\Gamma(\bf{E})$ and $\bf{E}$ share the same eigenvectors. Here we have:
\begin{equation}\label{25}
\Gamma({\bf{E}})=\sum_{i=1}^{N}\xi _{i}{\bf{u^{}_{li}u^{T}_{ri}}}
\end{equation}
In other words, the eigenvectors are fixed while the eigenvalues are free variables in this optimization. Therefore, it is easy to find the optimal solution:
\begin{equation}\label{26}
\xi_{i}=\sum_{j=1}^{N}c _{j}({\bf{v^{}_{lj}\cdot u^{}_{li}}})^{2}
\end{equation} where $(\cdot)$ denotes the inner product between two vectors. Then the expectation of $(v^{}_{lj}\cdot u^{}_{li})^{2}$ is considered as a more reasonable estimator:
\begin{equation}\label{27}
\xi_{i}=\sum_{j=1}^{N}c _{j}\mathbb{E}[{\bf{(v^{}_{lj}\cdot u^{}_{li}}})^{2}]
\end{equation} Although this framework only involves cleaning the eigenvalues, the eigenvectors are also thoroughly considered in this optimization problem, whose optimal solution (\ref{27}) can be regarded as converting the task of cleaning the eigenvectors into cleaning the eigenvalues. The $\mathbb{E}[{\bf{(v^{}_{lj}\cdot u^{}_{li}}})^{2}]$ can be analytically expressed in \cite{overlap}:
\begin{equation}\label{28}
\mathbb{E}[({\bf{v^{}_{lj}\cdot u^{}_{li}}})^{2}]=\frac{q\alpha(\lambda_{i})^{2}(\lambda_{i}\alpha(\lambda_{i})+c_{i})}
{\beta(\lambda_{i})^{2}+\gamma(\lambda_{i})^{2}}
\end{equation} where:
\begin{equation}\label{29}
\begin{split}
&\alpha(\lambda_{i})=(1-qh_{E}(\lambda_{i}))^{2}+q^{2}\pi^{2}\rho^{2}_{E}(\lambda_{i}) \\
&\beta(\lambda_{i})=(\lambda_{i}\alpha(\lambda_{i})+c_{i})(1-q)h_{E}(\lambda_{i})-\alpha(\lambda_{i})(1-q)
\\
&\gamma(\lambda_{i})=(\lambda_{i}\alpha(\lambda_{i})+c_{i})q\pi\rho_{E}(\lambda_{i})
\end{split}
\end{equation} where $h_{E}(\lambda_{i})$ is the real part of the Stieltjes transform (\ref{18}), and $\rho_{E}(\lambda_{i})$ is obtained from (\ref{19}).
\\\indent Combining (\ref{27}) and (\ref{28}), the cleaning function for the eigenvalues is obtained (derivations are shown in appendix. B):
\begin{equation}\label{30}
\xi_{i}=(1-qh_{E}(\lambda_{i}))(\lambda_{i}-(1-q)-2q\lambda_{i}h_{E}(\lambda_{i}))+q\varphi(\lambda_{i})
\end{equation} where:
\begin{equation}\label{30add}
\varphi(\lambda_{i})=1-h_{E}(\lambda_{i})(\lambda_{i}-(1-q))-q\lambda_{i}(\pi^{2}\rho^{2}_{E}(\lambda_{i})-h^{2}_{E}(\lambda_{i}))
\end{equation} By function (\ref{30add}), we can obtain the eigenvalues of the cleaned covariance matrix $\Gamma(\bf{E})$, then the cleaned eigenvalues of the matrix $\bf{F_{Z}}$ and $\bf{D_{Z}}$ are obtained successively by (\ref{15}). As same as reconstructing $\bf{C}$ from $\bf{E}$, we reconstruct $\bf{D_{H}}$ from $\bf{D_{Z}}$:
\begin{equation}\label{31}
{\Gamma\bf{(D_{Z}})}=\sum_{i=1}^{N+T}\xi_{Di}{\bf{u^{}_{lDi}}}{\bf{u}}_{{\bf{rDi}}}^{T}
\end{equation} where $\xi_{Di}$, $\bf{u^{}_{lDi}}$ and $\bf{u^{}_{rDi}}$ denote the estimated $i$-$th$ eigenvalue of $\bf{D_{H}}$ and the corresponding left and right eigenvector of $\bf{D_{Z}}$. Finally, the cleaned matrix $\Gamma(\bf{Z})$ is a part of $\Gamma(\bf{D_{Z}})$, following which we obtain the cleaned measurements $\gamma(\bf{z})$. The whole calculation flow of RBEC is shown in Fig. \ref{Fig.03}.
\\\indent \textit{Remark 1}: RBEC works in matrix-level, thus measurement noise in every entry of the matrix $\bf{Z}$ can be cleaned jointly. In this paper, historical data is used to form the matrix $\bf{Z}$, and is cleaned together with current data, so that it is an ensemble processing of measurement data. In this paper, for convenience, current measurements are solely extracted for the next stage.
\\\indent \textit{Remark 2}: Even if we reconstruct the Hermitian matrix $\bf{F_{Z}}$, we clean the eigenvalues of $\bf{E}$, because the R-transform of $\bf{F_{Z}-F_{H}}$ is not analytical.
\subsection{Two-stage State Estimation}
After RBEC, the cleaned measurement vector $\gamma(\bf{z})$ is input to WLS to estimate state variables. So it is a two-stage state estimation method, named RBEC-WLS (R-WLS), in which RBEC and WLS operate successively. The entire process of our method is shown in Algorithm. \ref{entire}.
\subsection{Boundary Condition}
Now we list the boundary condition of our matrix-level cleaning method. The first condition is that the size of $\bf{Z}$ is required to follow the Kolmogorov limit: (1) $N \rightarrow \infty$ and $T \rightarrow \infty$: the number of rows $N$ and columns $T$ should be sufficiently large. (2) $N\sim O(T) (N<T)$: $N$ is required to be comparable to $T$ \cite{free}. If $N>T$ or $T>>N$, the effect of our method will be unsatisfactory, which will be clearly discussed in case studies.
\\\indent The second boundary condition is that the variance of measurement error can be estimated. Since our method is based on the determinacy of the eigenvalues of large-dimensional Gaussian random matrices, measurement errors with different variance need to be normalized before operation.
\\\indent Also, other common settings of static state estimation are required, such as the information of topology and system parameters, Gaussian assumption of measurement error and so on.
\begin{algorithm}
\caption{The Entire Process of R-WLS}
\label{entire}
\begin{algorithmic}[1]
\State Normalize the measurement data by (\ref{7});
\label{a1}
\State Construct the matrix $\bf{Z}$ by (\ref{8});
\label{a2}
\State Construct the matrix $\bf{D_{Z}}$ by (\ref{11}) and $\bf{F_{Z}}$ by (\ref{12});
\label{a3}
\State Calculate the eigenvalues of $\bf{E}$;
\label{a4}
\State Calculate the eigenvectors of $\bf{D_{Z}}$;
\label{a5}
\State Clean the eigenvalues of $\bf{E}$ by function (\ref{30});
\label{a9}
\State Obtain the cleaned eigenvalues $\xi_{Di}$ of $\Gamma(\bf{D_{Z}})$ by (\ref{15});
\label{a10}
\State Reconstruct the matrix $\Gamma(\bf{D_{Z}})$ by (\ref{31});
\label{a11}
\State Choose the rows $1:N$ and columns $N+1:N+T$ in $\Gamma(\bf{D_{Z}})$ as $\Gamma(\bf{Z})$, thus the cleaned measurements are obtained;
\label{a12}
\State Input the cleaned measurement vector into WLS to estimate state variables.
\label{a13}
\end{algorithmic}
\end{algorithm}
\subsection{More Discussions}
The underlying mechanism inside our method is the well-known M-P law (see appendix. A). The M-P law reveals that the eigenvalues distribution of a Gaussian covariance matrix ${\bf{GG}}^{H}/T$ converges asymptotically to a deterministic probability distribution. The Stieltjes transform, R-transform and S-transform (see appendix. A) are ways by which the eigenvalues can be readily analyzed in a theoretical way. Specifically, the S-transform (\ref{S}) allows us to approximate the theoretical eigenvalues distribution of matrices product, such as ${\bf{GG}}^{H}$, ${\bf{HG}}^{H}$ and ${\bf{GH}}^{H}$. And the R-transform (\ref{20}) allows us to analytically compute the eigenvalues distribution of the sum of these matrices. So by these transforms, the eigenvalues of the population matrix $\bf{C}$ are connected with those of $\bf{E}$, and the overlaps of arbitrary eigenvectors of $\bf{E}$ and $\bf{C}$ are completely computable \cite{overlap}. Then an optimization scheme is designed to convert the task of cleaning the whole covariance matrix into cleaning its eigenvalues. The solution to this optimization problem merely involves the eigenvalues and the overlaps of its eigenvectors, which are completely available by these transforms, then the cleaning equation (\ref{30}) is obtained. In addition, this scheme tends to focus on cleaning the covariance matrix, so we propose a Hermitian matrix construction approach to enable the optimization scheme to clean the original measurements of power systems.
\begin{figure*}
    \subfigure[$\bf{E}$]{
    \begin{minipage}[t]{0.315\linewidth}
    \centering
    \includegraphics[height=1.6in,width=2.3in]{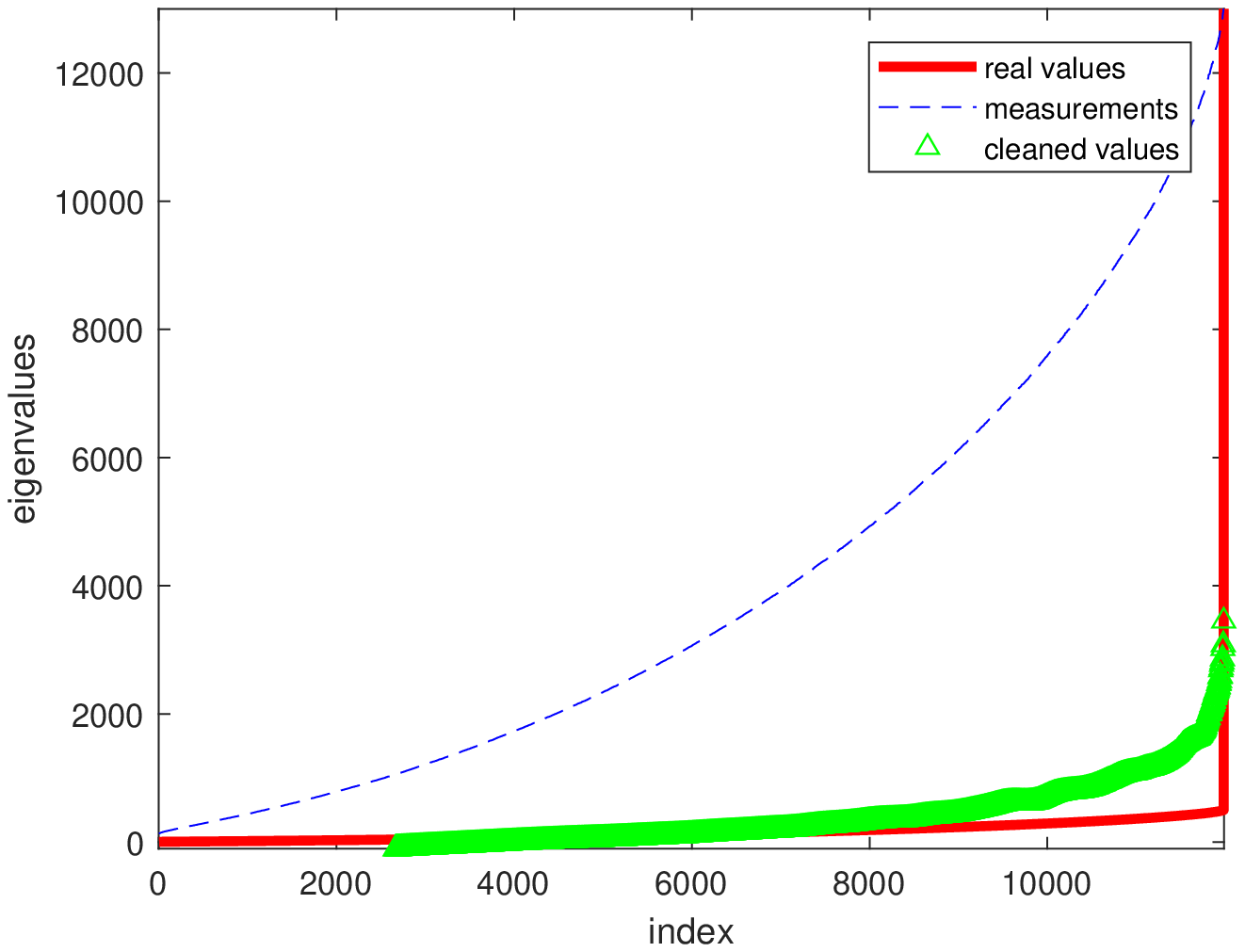}
    \label{Fig.05}\end{minipage}
    }
    \subfigure[$\bf{F_{Z}}$]{
    \begin{minipage}[t]{0.315\linewidth}
    \centering
    \includegraphics[height=1.6in,width=2.3in]{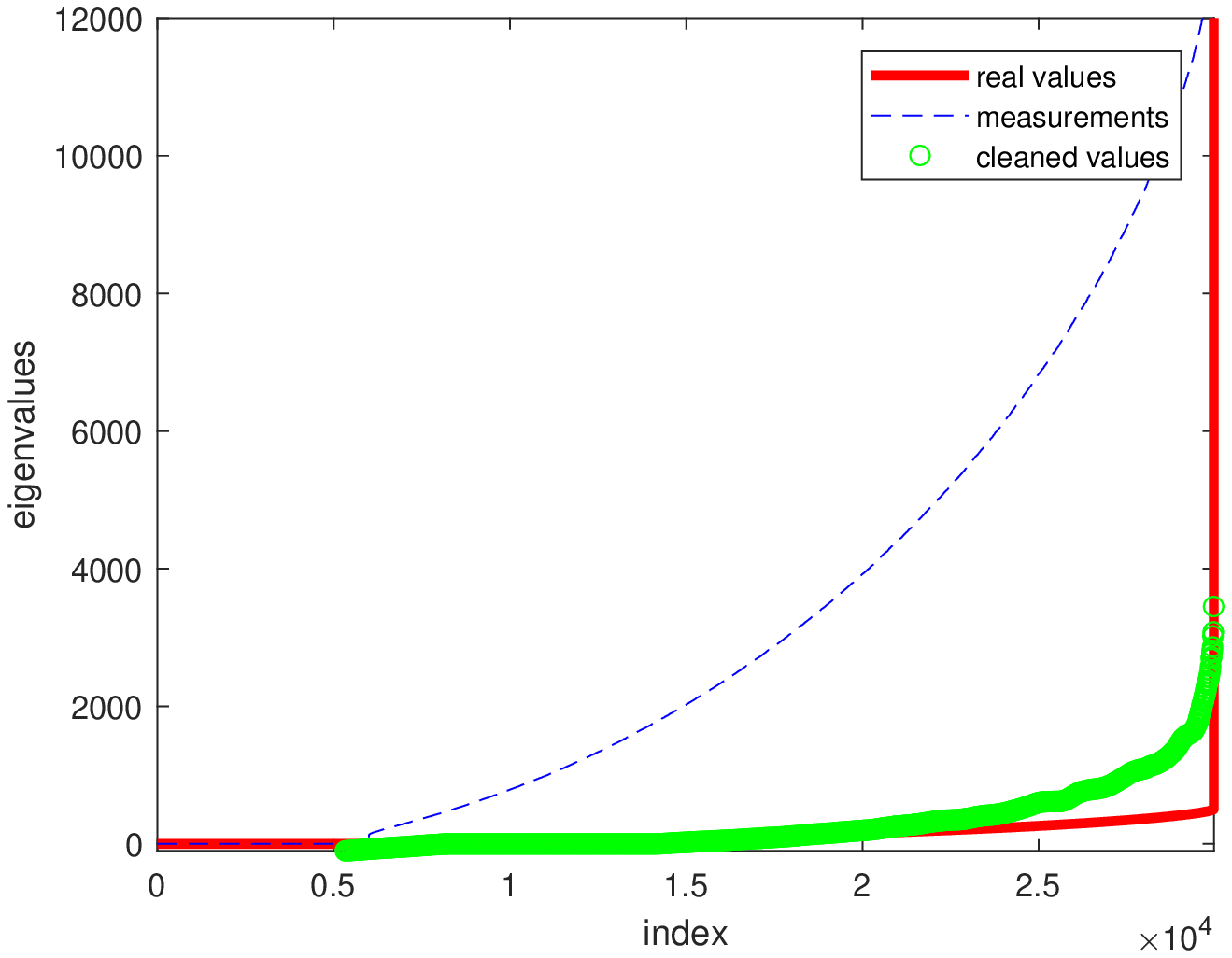}
    \label{Fig.06}\end{minipage}
    }
    \subfigure[$\bf{D_{Z}}$]{
    \begin{minipage}[t]{0.315\linewidth}
    \centering
    \includegraphics[height=1.6in,width=2.3in]{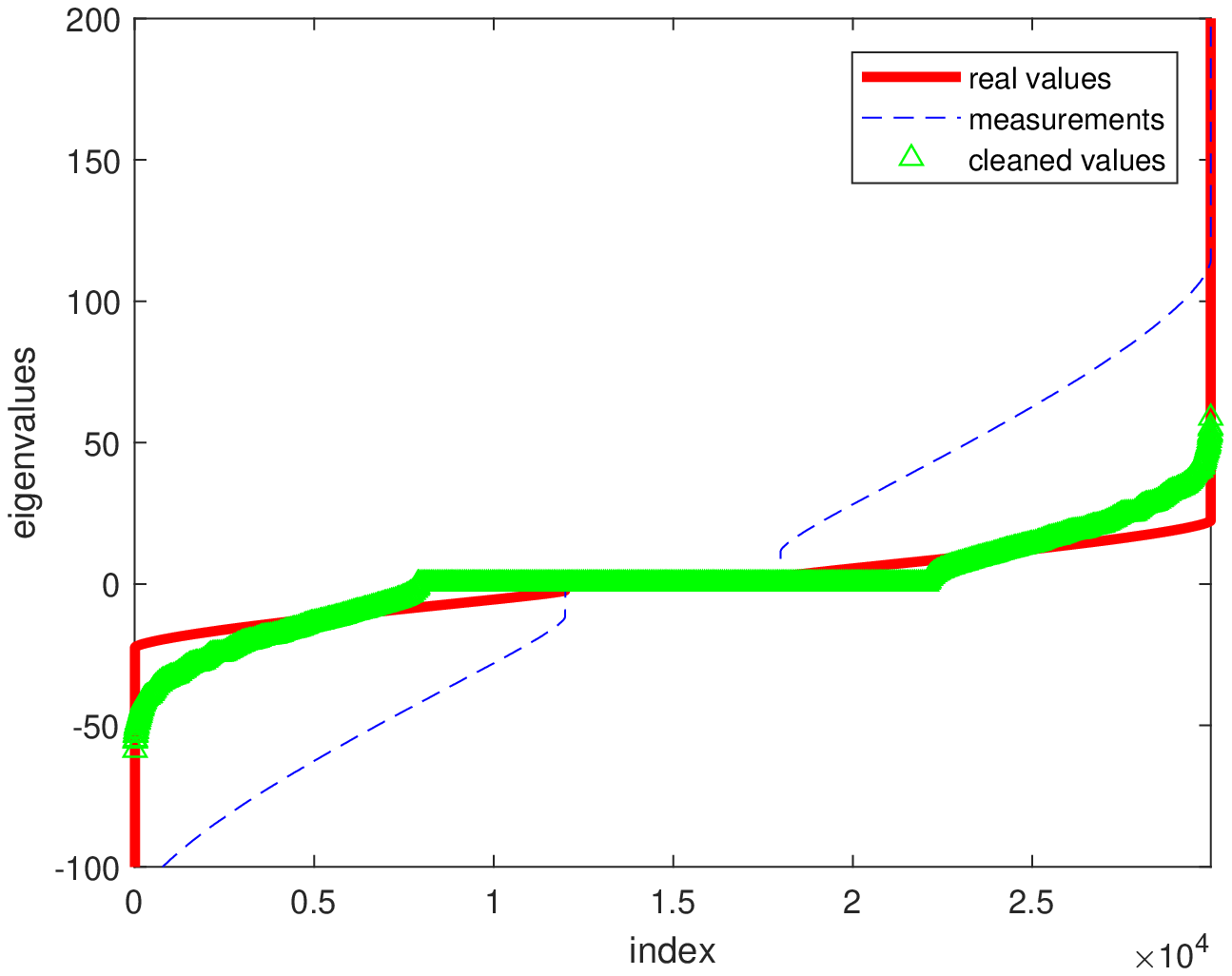}
    \label{Fig.07}\end{minipage}
    }
    \caption{The cleaning results of the eigenvalues of $\bf{E}$, $\bf{F_{Z}}$ and $\bf{D_{Z}}$ in case. 1. The abscissa is the numbering of the eigenvalues (sorted), as well as the ordinate is the eigenvalues.}\label{Fig.three}
\end{figure*}
\section{Case Study}
In this section, the accuracy and effectiveness of the proposed approach are verified explicitly. In the first case in European 1354-bus high voltage transmission system, the details of the calculation process are elaborated. In the second case, we test our approach using multiple testing systems and various magnitudes of measurement error. The third case shows the performance of our method in the case of different size ratios $q=T/N$ of the measurement matrix $\bf{Z}$. The fourth case is one in which the measured noise is modeled into non-Gaussian distributions. And the fifth case aims to demonstrate that our method is still valid when we need to divide the sample variables into matrices of appropriate size, in order to overcome the problem of too many measured variables. In the sixth case, other noise-cleaning methods are compared with our approach. The final case demonstrates that our method has little relevance to the operating state of power systems.
\subsection{Detailed Process}
This case is tested in European 1354-bus high voltage transmission system, which has 1354 bus, 1991 branches and 260 generators. In the simulation, Gaussian measurement errors with randomly selected bias $b_{i}\sim \mathcal{U}(-0.03, 0.03)$ are added, whose variance is 5\% of original values for power flows, as well as 1\% for voltage magnitudes. The initial values of voltage magnitudes for the iteration of WLS are randomly sampled from the Gaussian distribution $\mathcal{N}(1, 0.05)$, while the initial values of voltage angles are chosen from $\mathcal{N}(0, 0.157)$. The results are the average of ten identical experiments. The measurements from SCADA include branch active and reactive power from bus (1991+1991 variables), branch active and reactive power to bus (1991+1991 variables), nodal injective active and reactive power (1354+1354 variables) as well as nodal voltage magnitudes (1354 variables). Thus the size of the measurement vector is $12015\times 1$, then we choose $1.2*12015\approx 14418$ recent historical measurement vectors to construct a $12015\times 14418$ matrix.
\\\indent Using Algorithm. 1, the eigenvalues of $\bf{D_{Z}}$, $\bf{F_{Z}}$, $\bf{E}$ and the eigenvectors of $\bf{D_{Z}}$ can be calculated. Then we clean the eigenvalues of $\bf{E}$ by function (\ref{30}), followed by obtaining the cleaned eigenvalues of $\bf{F_{Z}}$ and $\bf{D_{Z}}$.
\\\indent The cleaned results of the eigenvalues of $\bf{E}$, $\bf{F_{Z}}$ and $\bf{D_{Z}}$ are shown in Fig. \ref{Fig.05}, Fig. \ref{Fig.06} and Fig. \ref{Fig.07} respectively. References \cite{lowrank1} and \cite{lowrank2} have demonstrated empirically that the measurement matrix in power systems is low-rank, which means that only a few eigenvalues (usually one or two) are far greater than zero, as the red lines reveal in Fig. $4\sim 6$. The blue dashed lines, denoting the eigenvalues of the measurement matrix in Fig. $4\sim 6$, are smoother curves (i.e., more eigenvalues are much greater than zero). Therefore, measurement noise breaks the low-rank property of the monitoring data matrix, yielding a lot of disturbing eigenvalues (components). And greatly reducing these disturbing eigenvalues in the absence of any knowledge of true values is the main task of RBEC, as the green lines reveal in Fig. $4\sim 6$. The explicit results of the cleaned eigenvalues are listed in Table. \ref{tab.eigen}. The mean absolute error (MAE) is used to measure the difference between two vectors:
\begin{equation}\label{37}
MAE=\frac{1}{N}\sum_{i=1}^{N}|\xi _{i}-\lambda_{i}|
\end{equation} where $\xi_{i}$ and $\lambda_{i}$ are the $i$-$th$ element in two vectors. The reason why the root mean square error (RMSE) is not selected is that the RMSE puts too much emphasis on big outliers. Hence if there are one or two values whose estimated errors are very large, squaring them will cause a very large RMSE, thus it cannot properly reflect the overall difference between two vectors. Furthermore, the mean absolute percent error (MAPE) is also dropped because it is severely influenced by small values. If there are one or two very small real values, the division w.r.t. them will result in a large MAPE even though the difference between real values and the estimation is not great.
\begin{table}
  \caption{The cleaning results of the eigenvalues of $\bf{E}$, $\bf{F_{Z}}$ and $\bf{D_{Z}}$. The error is measured by MAE}
  \centering
   \begin{tabular*}{6cm}{@{\extracolsep{\fill}}c|cc}
            \toprule[1.2pt]
            \multirow{1}*{Matrices} & Measurement error & Estimated error \\ \hline  \\
            $\bf{E}$ & 1623.3825 & 102.3944 \\
            $\bf{F_{Z}}$ & 1448.2478 & 93.0455 \\
            $\bf{D_{Z}}$ & 62.4250 & 4.0329
            \\\toprule[1.2pt]
        \end{tabular*}\label{tab.eigen}
\end{table}
\\\indent As shown in Table. \ref{tab.eigen}, the errors of the eigenvalues of $\bf{E}$, $\bf{F_{Z}}$ and $\bf{D_{Z}}$ are greatly reduced. Then we construct the matrix $\Gamma(\bf{D_{Z}})$ by function (\ref{31}), followed by obtaining $\Gamma(\bf{Z})$ and $\gamma(\bf{z})$. Compared with the measurement vector $\bf{z}$ with the MAE of $0.0395$, the MAE of the cleaned vector $\gamma(\bf{z})$ is only $0.0082$ left, proving that our RBEC is effective to clean measurement error.
\\\indent The MAE of the state vector decreases from $0.0183$ to $0.0011$, demonstrating that our method significantly improves the classic WLS based state estimation.
\\\indent Then the estimation of measured variables is calculated by $\bf{\hat{z}=h(\hat{x})}$. The estimation error of the measurement vector is shown in Table. \ref{tab.measring}, where Pt, Pf, Pb, Qt, Qf, Qb, Vm denote active power to bus, active power from bus, nodal injective active power, reactive power to bus, reactive power from bus, nodal injective reactive power and voltage magnitudes, respectively. We must express the point that the estimation of measured variables $\bf{\hat{z}}$ is not the cleaned measurement vector $\gamma(\bf{z})$ which is obtained after WLS. Although the MAE of $\gamma(\bf{z})$ is smaller than $\bf{\hat{z}}$ in this case, $\bf{\hat{z}}$ is a more reasonable estimation result, for $\bf{\hat{z}}$ is more in line with the physical equations of the testing system.
\begin{table}
  \caption{The MAE of the estimation error of measured variables (Units: p.u.).}
  \centering
   \begin{tabular*}{6.4cm}{@{\extracolsep{\fill}}|c|c|c|c|}
            \hline
            \multirow{1}*{Variables} & Measurements & WLS & R-WLS  \\ \hline
            Pt & 0.0440 & 3.5306 & 0.1206  \\  \hline
            Pf & 0.0469 & 3.6494 & 0.1315  \\  \hline
            Pb & 0.0390 & 2.9567 & 0.1287  \\   \hline
            Qt & 0.0441 & 3.6455 & 0.1360  \\   \hline
            Qf & 0.0473 & 3.9787 & 0.1324  \\   \hline
            Qb & 0.0350 & 2.4066 & 0.0978  \\   \hline
            Vm & 0.0062 & 0.0039 & 0.0012  \\   \hline
        \end{tabular*}\label{tab.measring}
\end{table}
\subsubsection{Problem of Residual}
\begin{figure}
  \centering
  \includegraphics[height=2.3in,width=3.3in]{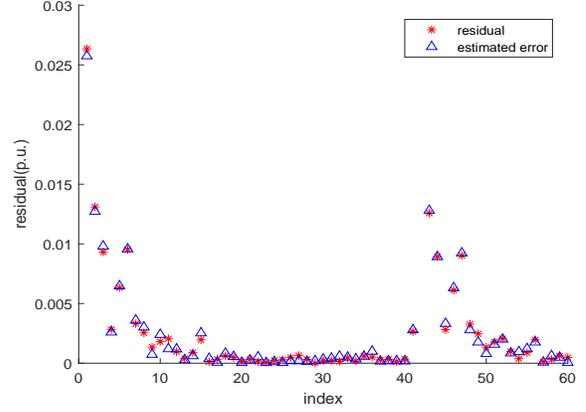}\\
  \caption{The comparison between the residual and the estimated error after R-WLS. The settings of the simulation are the same as Fig. \ref{Fig.01}.}\label{Fig.09}
\end{figure}

Our original intention of designing RBEC is to handle the problem of WLS that the residual is only a part of estimated error. As shown in Fig. \ref{Fig.01}, partial estimated errors are also significant when most of residuals are shrunk to small values. To prove that our R-WLS is able to overcome this problem, we employ the same simulation settings with the quick fact in IEEE 30-bus system. As shown in Fig. \ref{Fig.09}, both residuals and estimation errors become very small.
\subsection{Numerous Tests}
The above case clarifies the detailed process of our approach. Now, in this case, numerous tests in different power systems are conducted using different magnitudes of measurement error. The size ratio $q$ is set as $1.2$ in all tests. The initial values of voltage magnitudes are randomly sampled from the Gaussian distribution $\mathcal{N}(1, 0.05)$, while the initial values of voltage angles are chosen from $\mathcal{N}(0, 0.157)$. The IEEE 30-bus, 57-bus, 118-bus, 300-bus systems, European 1354-bus high voltage transmission system, Polish 3120-bus system at summer 2008 morning peak, French 6468-bus very high voltage and high voltage transmission network, and European 9241-bus system are utilized to test our method. The explicit parameters of these systems are referred to \cite{1354pegase}. In terms of measurement errors, 2.5\%, 5\%, 7.5\%, 10\% error for power flows and 0.5\%, 1\%, 1.5\%, 2\% for voltage magnitudes are set. The sizes of the matrices are listed in Table. \ref{tab.size}, where the number of rows corresponds to the number of sample variables, and the number of columns represents the length of the split window. The explicit results are shown in Table. \ref{tab.numerous}.
\begin{table}
  \caption{The matrix sizes of different systems.}
  \centering
   \begin{tabular*}{5cm}{@{\extracolsep{\fill}}c|c}
            \toprule[1.2pt]
            \multirow{1}*{Systems} & Sizes of matrices \\ \hline  \\
            European 1354-bus & 12015$\times$14418 \\
            Polish 3120-bus & 24117$\times$28940 \\
            French 6468-bus & 55389$\times$66467 \\
            European 9241-bus & 91904$\times$110285
            \\\toprule[1.2pt]
        \end{tabular*}\label{tab.size}
\end{table}
\begin{table*}
  \caption{The MAE of cases of different systems and various magnitudes of measurement error (Units: p.u.)}
  \centering
   \begin{tabular*}{14cm}{@{\extracolsep{\fill}}|c|c|c|c|c|c|c|c|c|}
            \hline
            \multirow{1}*{Error of power flows} & \multicolumn{2}{c|}{2.5\%} & \multicolumn{2}{c|}{5\%} & \multicolumn{2}{c|}{7.5\%} & \multicolumn{2}{c|}{10\%} \\ \hline
            \multirow{1}*{Error of Vm} & \multicolumn{2}{c|}{0.5\%} & \multicolumn{2}{c|}{1\%} & \multicolumn{2}{c|}{1.5\%} & \multicolumn{2}{c|}{2\%} \\ \hline
            Methods & WLS & R-WLS & WLS & R-WLS & WLS & R-WLS & WLS & R-WLS  \\ \hline
            IEEE 30-bus & 0.0039 & 0.0010 & 0.0082 & 0.0015 & 0.0154 & 0.0015 & 0.0174 & 0.0014   \\
            IEEE 57-bus & 0.0049 & 0.0010 & 0.0108 & 0.0016 & 0.0166 & 0.0021 & 0.0195 & 0.0022  \\
            IEEE 118-bus & 0.0052 & 0.0007 & 0.0146 & 0.0013 & 0.0186 & 0.0017 & 0.0174 & 0.0018   \\
            IEEE 300-bus & 0.0082 & 0.0006 & 0.0184 & 0.0008 & 0.0233 & 0.0014 & 0.0239 & 0.0017   \\
            European 1354-bus & 0.0063 & 0.0011 & 0.0183 & 0.0011 & 0.0245 & 0.0014 & 0.0273 & 0.0019   \\
            Polish 3120-bus & 0.0332 & 0.0013 & 0.0482 & 0.0015 & 0.0547 & 0.0015 & 0.0674 & 0.0018   \\
            French 6468-bus & 0.0186 & 0.0010 & 0.0315 & 0.0013 & 0.0424 & 0.0015 & 0.0470 & 0.0018  \\
            European 9241-bus & 0.0271 & 0.0005 & 0.0494 & 0.0007 & 0.0631 & 0.0009 & 0.0780 & 0.0013   \\
            \hline
        \end{tabular*}\label{tab.numerous}
\end{table*}
\\\indent According to Table. \ref{tab.numerous}, two important properties of our method are clearly revealed. At first, with the increase of system scale, the performance of our approach improves, while the performance of traditional WLS becomes worse by contrast. This can be explained by that some equations of RMT on which our method based asymptotically hold with the matrix size increasing to infinity, such as the inverse stieltjes transform (\ref{19}) and the R-transform of Gaussian covariance matrices. RMT derives these equations by assuming that the matrix size converges to the Kolmogorov limit. So that a larger matrix would lead to a better result which is closer to the theoretical conclusion. Therefore, our approach is particularly suitable for large interconnected systems.
\\\indent Secondly, the results are getting worse as a response to the increase of measurement error. Obviously, the greater error will make it more difficult for WLS to find the optimal solution, as well as for our RBEC to clean measurements. The numerous tests, using different systems and various magnitudes of measurement error, demonstrate the effectiveness and generality of our method.
\subsection{Case with Different Size Ratios}
\begin{figure}
  \centering
  \subfigure[$q$]{
    \begin{minipage}{0.45\linewidth}
    \centering
    \includegraphics[height=1.3in,width=1.8in]{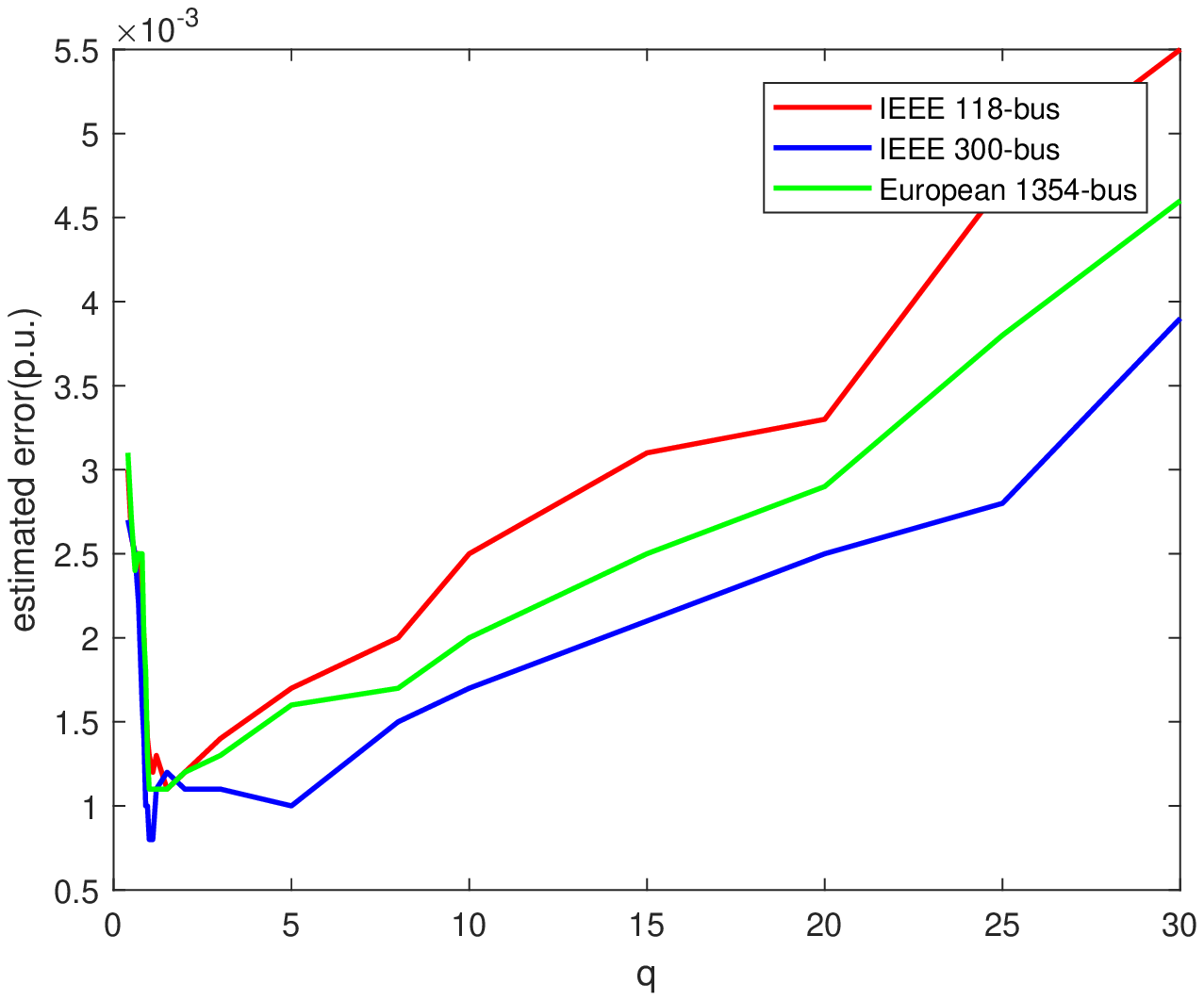}
    \label{Fig.101}\end{minipage}
    }
  \subfigure[$ln(q)$]{
    \begin{minipage}{0.48\linewidth}
    \centering
    \includegraphics[height=1.3in,width=1.8in]{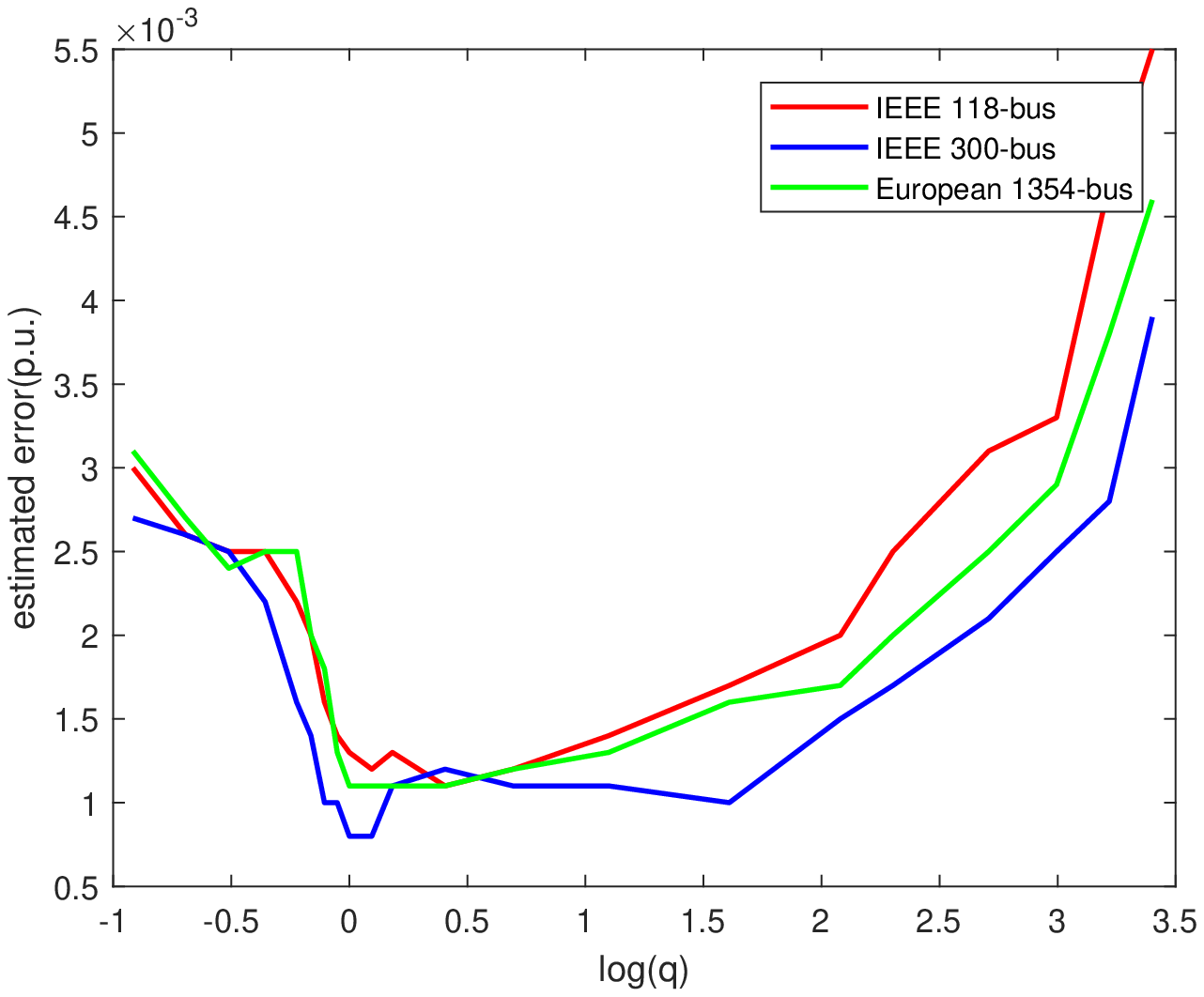}
    \label{Fig.102}\end{minipage}
    }
  \caption{The results of different $q$ in IEEE 118-bus, 300-bus and European 1354-bus systems. To clearly reveal the results when $q<1$, we modify the abscissa as $ln(q)$ in subfigure(b). }\label{Fig.10}
\end{figure}

As introduced above, the size of the measurement matrix $Z$ is required that the number of rows $N$ should be comparable to the number of columns $T$, i.e., $N\sim O(T) (N<T)$ or $q=T/N>1$. So it is very meaningful to test our approach of different $q$ values, if or if not the requirements of size are met. In this simulation, $q$ is set to range from 0.3 to 50, and other settings are the same as the first case. The results are shown in Fig. \ref{Fig.10}.
\\\indent According to Fig. \ref{Fig.10}, if $T>>N$, the effect of our RBEC becomes unpleasant, since many analytical equations of RMT, such as the M-P Law (\ref{36}) and the inverse Stieltjes transform (\ref{19}), do not hold. Refer to appendix. D. for more information about what role this condition plays. Additionally, when $N<T$, the effect is also unacceptable because the $N$-dimensional noise space of measurement error cannot be reconstituted. Specifically, reconsidering model (\ref{10}) and (\ref{bujiao}), the noise space is $N$-dimensional since the stochastic measurement error of every sensor or substation (every row) is independent. However, the number of non-zero eigenvalues is $\min \{T,N\}$. So if $T<N$, RBEC can only span a $T$-dimensional subspace instead of the entire noise space, which is not conducive to our approach.
\\\indent Roughly speaking, the results are satisfactory and keep steady when $q$ ranges approximately from 1 to 8. And the larger the matrix $\bf{Z}$, the more computing resources our approach will consume. Therefore, the most reasonable value of $q$ is little greater than 1.
\subsection{Case with Different Noise Models}
This subsection discusses the effectiveness of our method when measurement error is not Gaussian distributed. The mathematical derivation of our method is based on the assumption that measurement error follows a Gaussian distribution, for Gaussian noise is the commonest model of unknown measurement error \cite{measurement}. However, some unclear randomness may violate this assumption, leading to non-Gaussian noise. So it is necessary to discuss the performance of our approach in the case of non-Gaussian distributions of measurement error.
\\\indent In this case, the Laplace distribution, semi-circle distribution (SC), symmetric-linear error (SL) and normal-inverse Gaussian distribution (NIG) are simulated, for they are somewhat similar to noise models. The probability density functions (p.d.f.) and coefficients are listed in Table. \ref{tab.func}. The settings of coefficients aim to make the variance of these distributions equal to $\sigma_{i}$. The Monte Carlo method is used to generate samplings of these distributions. This case uses European 1354-bus system, and the averaged results of five tests are shown in Table. \ref{tab.funresult}. The Inc.Rat in Table. \ref{tab.funresult} means the increasing ratio of MAE, i.e., $Inc. Rat=(MAE_{W}-MAE_{R})/MAE_{W}$.
\\\indent The results agree with our analysis that in the case of Gaussian noise, the promotion of traditional WLS is the highest. And then, though the MAE of WLS in the case of NIG distribution $0.0134$ is smaller than the Gaussian case $0.0164$, the MAE of R-WLS in the NIG case $0.0026$ is not as small as the Gaussian case $0.0013$, which reveals that the cleaning effect of NIG noise is not as high as the case of Gaussian error. Furthermore, in all cases, R-WLS successfully improves the performance of WLS, showing that our approach has certain generality and can be effective in practice.
\begin{table}
  \caption{The P.d.f and coefficients of various noise models.}
  \centering
   \begin{tabular*}{8cm}{@{\extracolsep{\fill}}ccc}
            \toprule[1.2pt]
            \multirow{1}*{Distributions} & P.d.f & Coefficients \\ \hline  \\
            Laplace  & $f(x)=\frac{1}{2b}exp(-\frac{|x-\mu |}{b})$ & $\mu=0$,$b=\sigma_{i}/\sqrt{2}$ \\
            SC  & $f(x)=\frac{2}{\pi a^{2}}\sqrt{a^{2}-x^{2}}$ & $a=2\sigma_{i}$ \\
            SL & $y=\frac{1}{a^{2}}(a-|x|)$ & $a=\sqrt{6}\sigma_{i}$ \\
            NIG & $f(x)=NIG(x)$ & see appendix. C\\
            Gaussian & $f(x)=\frac{1}{\sqrt{2\pi }b }exp(-\frac{(x-\mu )^{2}}{2b ^{2}})$ & $\mu=0$,$b=\sigma_{i}$ \\
            \\\toprule[1.2pt]
        \end{tabular*}\label{tab.func}
\end{table}
\begin{table}
  \caption{The MAE of the estimation error in the case of different noise distributions (MAE, Units: p.u.).}
  \centering
   \begin{tabular*}{5.9cm}{@{\extracolsep{\fill}}|c|c|c|c|}
            \hline
            \multirow{1}*{Distributions} & WLS & R-WLS & Inc.Rat \\ \hline
            Laplace & 0.0246 & 0.0102 & 58.5\% \\  \hline
            SC & 0.0427 & 0.0292 & 31.6\% \\  \hline
            SL & 0.0280 & 0.0141 & 49.7\% \\   \hline
            NIG & 0.0195 & 0.0032 & 83.6\% \\   \hline
            Gaussian & 0.0183 & 0.0011 & 94.0\% \\   \hline
        \end{tabular*}\label{tab.funresult}
\end{table}
\subsection{Case with Divided Matrices}
According to the boundary condition introduced in Section~\Rmnum{3}, the size of $\bf{Z}$ is constrained by $N<T$. In practice, the number of measured variables can easily be tens of thousands, so the measurement samples should span a time window of commensurable size. This may take a long time during which the system topology may have changed. To address this problem, we divide measured variables into sets of appropriate size if there are too many measured variables, or the equipment for running algorithms is limited. For the sake of brevity, this case uses European 1354-bus high voltage transmission network, which contains 12015 measured variables, in order are active power to bus Pt, active power from bus Pf, injective active power Pb, reactive power to bus Qt, reactive power from bus Qf, injective reactive power Qb, nodal voltage magnitudes Vm. We divide them into seven groups according to their physical meaning. After cleaning all respective matrices, we gather them to operate WLS.
\\\indent The explicit sizes of matrices and the results of cleaning are shown in Table. \ref{tab.division}. After cleaning, the averaged MAE of the measured vector decreases from 0.0395 to 0.0097. The MAE of the state vector after R-WLS is 0.0014, while MAE is 0.0011 if measured variables are cleaned in one matrix. Therefore, by dividing the variables, we can speed up the operation without much worsening of effectiveness.
\begin{table}
  \caption{The cleaning effects of divided matrices (MAE, Units: p.u.).}
  \centering
   \begin{tabular*}{6.8cm}{@{\extracolsep{\fill}}|c|c|c|c|}
            \hline
            \multirow{1}*{Variables} & Size & Measurement & Cleaned  \\ \hline
            Pt & $1991\times2389$ & 0.0440 & 0.0117  \\  \hline
            Pf & $1991\times2389$ & 0.0469 & 0.0115  \\  \hline
            Pb & $1354\times1625$ & 0.0390 & 0.0106 \\   \hline
            Qt & $1991\times2389$ & 0.0441 & 0.0102  \\   \hline
            Qf & $1991\times2389$ & 0.0473 & 0.0146  \\   \hline
            Qb & $1354\times1625$ & 0.0350 & 0.0083  \\   \hline
            Vm & $1354\times1625$ & 0.0062 & 0.0016  \\   \hline
        \end{tabular*}\label{tab.division}
\end{table}
\subsection{Comparison of Different Cleaning Methods}
It is necessary to compare our approach with other RMT based cleaning methods. In this case, the nonlinear shrinkage (NLS) \cite{NLS}, loo-cross-validation covariance (CVC) \cite{CVC} and isotonic regression based CVC (iso-CVC) \cite{CVC} are compared. The simulation settings are the same as those in Section. \Rmnum{4}. B, and the results are shown in Table. \ref{tab.CVC}.
\\\indent According to Table. \ref{tab.CVC}, the comparison results in power systems of different sizes are different. In IEEE 30-bus system, our method is superior to the other methods, which shows that our method has weaker requirement for matrix size ($N$ and $T$ should be large enough). This requirement for matrix size, or more broadly the Kolmogorov limit, is general and fundamental for all RMT based methods. So it can be seen that our method is more tolerant for this crucial condition. As for the case of IEEE 57-bus, 118-bus and 300-bus systems, our method also obtains comparable results, demonstrating the effectiveness and competitiveness of our approach.
\begin{table}
  \caption{The Comparison results of different cleaning methods (MAE, Units: p.u.)}
  \centering
   \begin{tabular*}{8.4cm}{@{\extracolsep{\fill}}|c|c|c|c|c|c|}
            \hline
            Methods & WLS & RBEC & NLS & CVC & iso-CVC \\ \hline
            IEEE 30-bus & 0.0082 & 0.0015 & 0.0019 & 0.0074 & 0.0016   \\
            IEEE 57-bus & 0.0108 & 0.0016 & 0.0019 & 0.0089 & 0.0017    \\
            IEEE 118-bus & 0.0146 & 0.0013 & 0.0015 & 0.0102 & 0.0012   \\
            IEEE 300-bus & 0.0184 & 0.0008 & 0.0012 & 0.0098 & 0.0007  \\
            \hline
        \end{tabular*}\label{tab.CVC}
\end{table}
\subsection{Case with time-varying operating state}
\begin{figure}
  \centering
  \subfigure[load settings]{
    \begin{minipage}{0.45\linewidth}
    \centering
    \includegraphics[height=1.3in,width=1.8in]{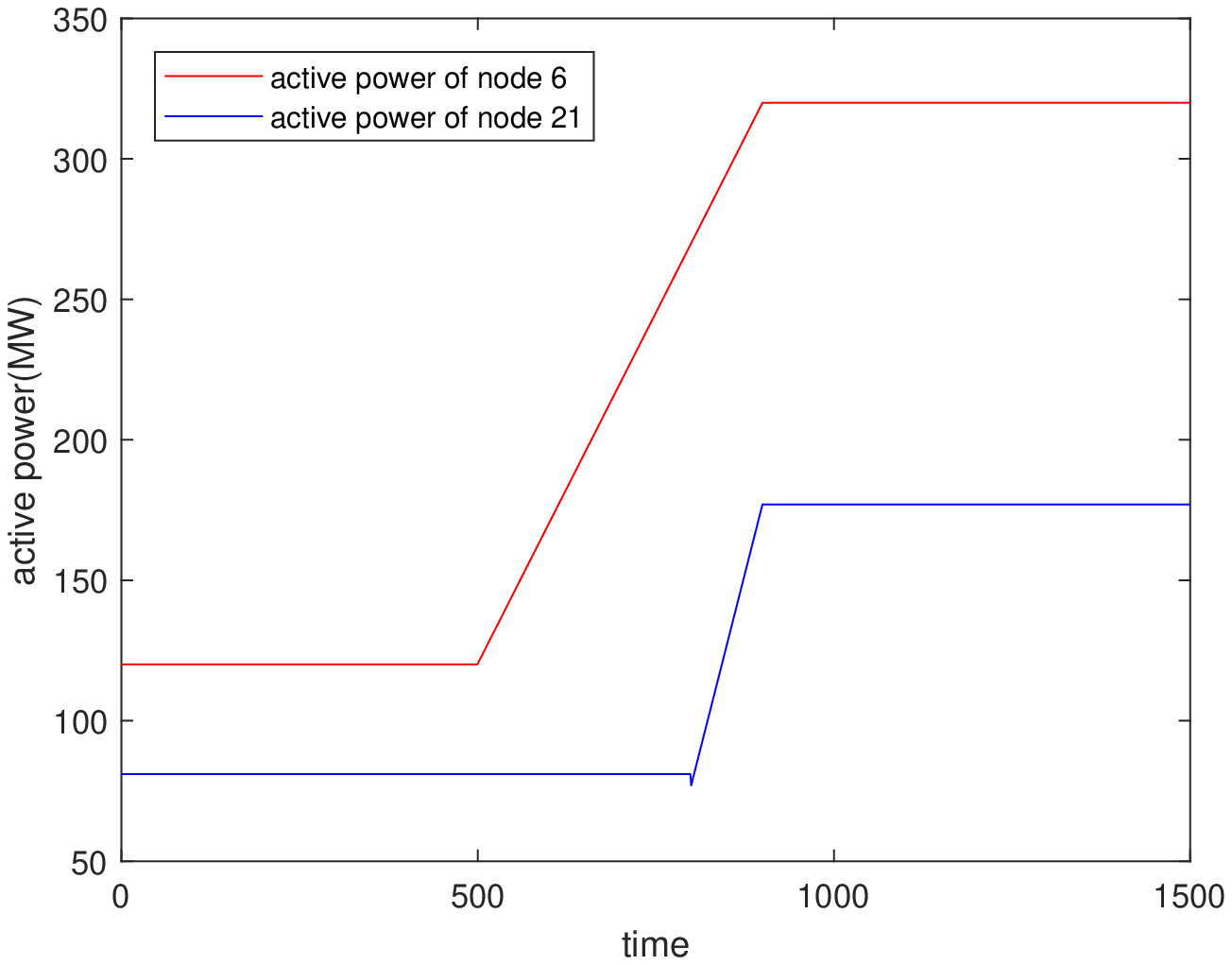}
    \label{Fig.tv1}\end{minipage}
    }
  \subfigure[MAE]{
    \begin{minipage}{0.48\linewidth}
    \centering
    \includegraphics[height=1.3in,width=1.8in]{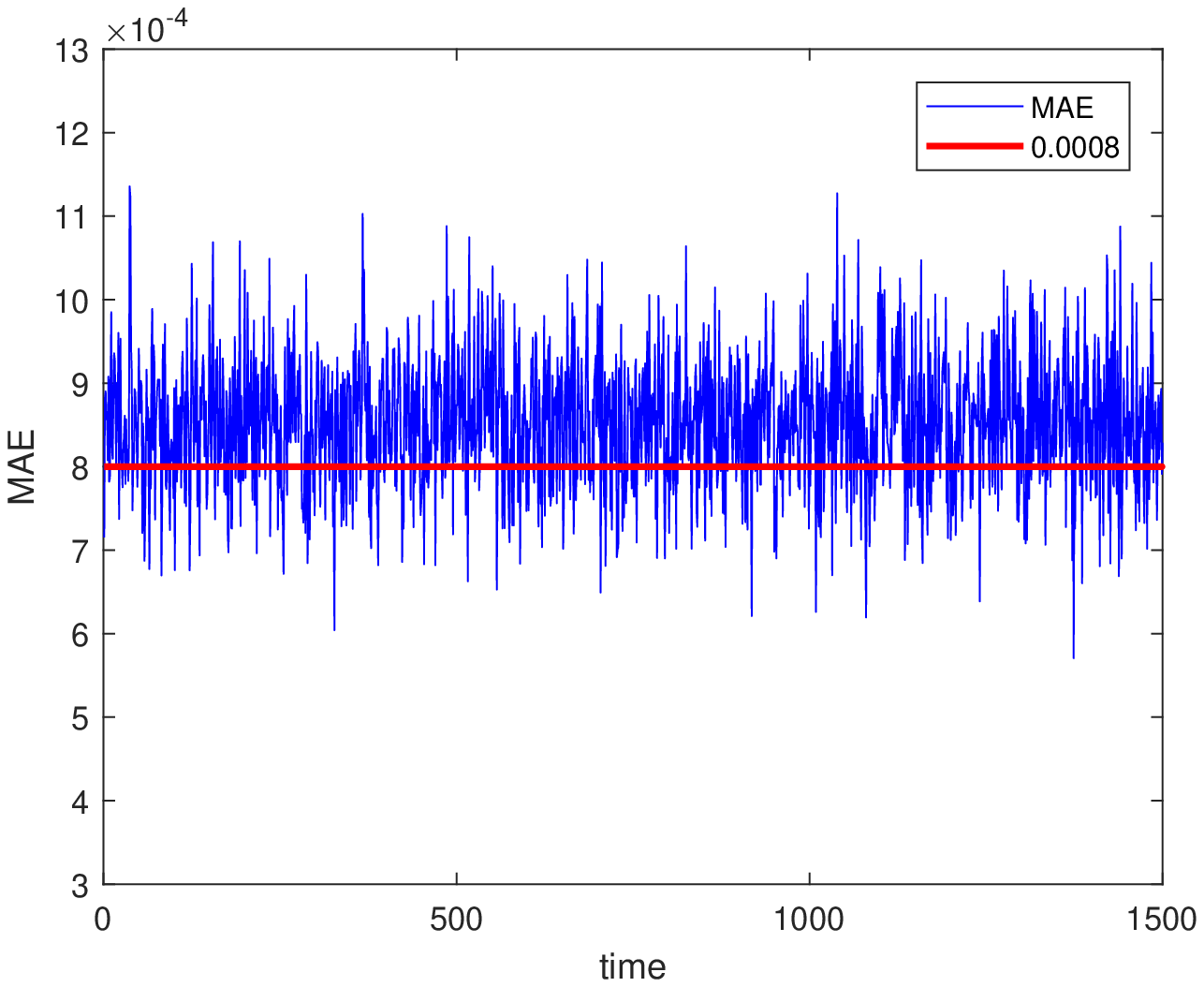}
    \label{Fig.tv2}\end{minipage}
    }
  \caption{The changing profile of active loads and the MAE over time.}\label{Fig.tv}
\end{figure}

In this case, we aim to illustrate the performance of the proposed method under the scenario of time-varying operating state. We use IEEE 300-bus system, and the same settings of measurement error and initialization method are adopted. The active load of node 6 and node 21 is selected to be time-varying, while the others remain constant. The details of the load settings are shown in Fig.~\ref{Fig.tv}(a).
\\\indent The MAE curve over time is shown in Fig.~\ref{Fig.tv}(b). According to the results, the effectiveness of our approach remains almost independent of the time-varying operating state. This is because our cleaning method rebuilds the measurement model in matrix-level, and it is based on the determinacy of the eigenvalues of $\bf{G}$ in the limit of large dimension. Therefore, whatever the matrix $\bf{H}$ is, as long as $\bf{H}$ follows the Kolmogorov limit, our method is capable of working normally. Besides, the results fluctuate from around 0.00075 to 0.00095, because the measurement noise imposed in each experiment is randomly generated.
\subsection{Case of Inaccurate Variance}
\begin{figure}
  \centering
  \includegraphics[height=2.2in,width=3.2in]{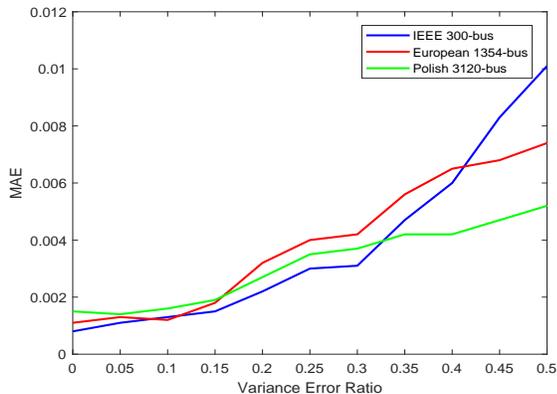}\\
  \caption{The results of different error ratios of variance estimation.}\label{Fig.last}
\end{figure}

As shown in the second assumption, the variance of measurement error is assumed to be known. However, in practice, the variance estimation may fail to obtain accurate results, due to the lack of historical data, sudden events and some systematic errors. Thus in this case, we examine the performance of our approach when the estimated variance of measurement error is also biased from the accurate variance. We use European 1354-bus, Polish 3120-bus and IEEE 300-bus system, and adopt the same hyper parameters and simulation settings with the above cases. The error ratios of variance estimation range from 0 to 50\% of the real variance. The testing results are shown in Fig. \ref{Fig.last}.
\\\indent From the results, the influence of inaccurate variance estimation decreases with the system scale increasing, especially when the variance error ratio is great. Besides, even though the inaccurate variance estimation generally reduces the effectiveness of our method, the MAE of final results is still better than those without cleaning.

\section{Conclusion}
In this paper, a new opinion was put forward on the key problem of state estimation, that is, estimated error comes from the residual and measurement error, but traditional WLS does not take the effect of measurement error into account. Therefore, we proposed a data-driven method to overcome this problem by cleaning measurement error at first. Combined with WLS, a two-stage state estimation model was conducted. Our method is based on the deterministic property about eigenvalues distribution of the fully stochastic measurement error in matrix-level, and it is the first time to use RMT based noise-cleaning method to improve state estimation. Additionally, another innovation is the Hermitian matrix construction, which is a kind of extension of previous cleaning models for covariance matrices, so that they can be applied to clean measurement matrices of power systems. Our method not only has strict mathematical deductions and precise theoretical supports, but also performs well in practical applications. The numerous tests, using different testing systems and various hyper-parameters, proved the effectiveness and advantages of our method. In the future, we will attempt to clean measurement noise in the iterative process of WLS.
\begin{appendices}
\section{Random Matrices Basics}
\textbf{Theorem 1: M-P Law} For an $N\times T$ matrix $\bf{G}$ whose entries follow an identical and independent Gaussian distribution $\mathcal{N}(0,\sigma^{2})$, $N\rightarrow \infty $, $T\rightarrow \infty$ and $T\sim O(N)$. Then the spectrum distribution of its covariance matrix ${\bf{GG}}^{T}/T$ asymptotically follows the M-P distribution \cite{mp}:
\begin{equation}\label{36}
\lambda\sim \frac{1}{2\pi \lambda q\sigma ^{2}}\sqrt{(b-\lambda)(\lambda-a)}
\end{equation} where $a<\lambda <b$, $q=T/N$, $a=\sigma ^{2}(1-\sqrt{q})^{2}$ and $b=\sigma ^{2}(1+\sqrt{q})^{2}$.
\\\indent \textbf{Theorem 2: Stieltjes Transform} One of the most general resolvent of the covariance matrix $\bf{E}$ is:
\begin{equation}\label{17}
G_{E}(s)=(sI_{N}-{\bf{E}})^{-1}
\end{equation} where $s\in \mathbb{C^{+}}$ is a complex variable, and ${\bf{I_{N}}}\in \mathbb{R}^{N\times N}$ is the unit matrix. The Stieltjes transform is defined as \cite{free}:
\begin{equation}\label{18}
g_{E}(s)=\frac{1}{N}Tr({G_{E}}(s))=\frac{1}{N}\sum_{i=1}^{N}\frac{1}{s-\lambda _{i}}
\end{equation} where $Tr(\cdot )$ is the trace of $\bf{E}$. $N$ denotes the number of rows of $\bf{E}$, while $\lambda_{i}$ is the $i$-$th$ eigenvalue of $\bf{E}$.
\\\indent \textbf{Theorem 3: Inverse Stieltjes Transform} The inverse transform of the Stieltjes transform is;
\begin{equation}\label{19}
\rho_{E}(\lambda _{i})=\frac{1}{\pi }Im[\lim_{\eta \rightarrow 0}g_{E}(\lambda _{i}-i\eta)] \end{equation} where $\eta$ is a small integer closed to zero, $Im[\cdot]$ denotes the image part.
\\\indent \textbf{Theorem 4: R-Transform} The R-transform is defined as:
\begin{equation}\label{20}
R_{E}(s)=<g_{E}(s)>^{-1}-\frac{1}{s}
\end{equation} where $<\cdot>^{-1}$ represents the inverse function. The R-transform of a sum of matrices is equal to the sum of their respective R-transforms \cite{free}:
\begin{equation}\label{21}
R_{A+B}(s)=R_{A}(s)+R_{B}(s)
\end{equation} The R-transform is a great tool to analyze additive Gaussian errors, since by the R-transform we can obtain the spectrum distribution of $\bf{A+B}$ from $\bf{A}$ and $\bf{B}$.
\\\indent \textbf{Theorem 5: S-Transform} The S-transform allows us to compute the eigenvalues distribution of matrices product. The S-transform is defined as:
\begin{equation}\label{S}
S_{E}(s)=\frac{s+1}{s\Gamma^{-1}_{s}}
\end{equation} where $\Gamma(s)=sg_{E}(s)-1$. The S-transform of a product of matrices is identical to the product of their respective S-transforms:
\begin{equation}\label{Smul}
S_{AB}(s)=S_{A}(s)S_{B}(s)
\end{equation} From the above basic equations, the eigenvalues distribution of a sum or a product of different or identical matrices can be fully computable. So reconsidering the model (\ref{bujiao}), the eigenvalues of each term can be analytically connected.
\section{Proof of Function (\ref{30})}
According to function (\ref{27}) and (\ref{19}), and the equation $G_{E}(z)=(z(1-qg_{E}(z))-(1-q)-(1-qg_{E}(z))^{-1}{\bf{C}})^{-1}$, we have:
\begin{equation}\label{39}
\begin{split}
\xi_{i}
& =\sum_{j=1}^{N}c_{j}\mathbb{E}[({\bf{v^{}_{lj}\cdot u^{}_{li}}})^{2}]\\
& =\int c\rho _{c}N\mathbb{E}[{\bf{(v^{}_{lj}\cdot u^{}_{li}}})^{2}]dc\\
& =\frac{1}{\pi \rho _{E}(\lambda _{i})}\lim_{z\rightarrow \lambda _{i}- \eta }Im[\int \frac{c\rho _{C}(c)}{1-qg_{E}(s)-c}dc]\\
&=\frac{1}{\pi \rho _{E}(\lambda _{i})}\lim_{s\rightarrow \lambda _{i}- \eta }Im[ Tr(G_{E}(s){\bf{C}})]\\
\end{split}
\end{equation}
And the right part is:
\begin{equation}\label{40}
\begin{split}
Tr(G_{E}(s){\bf{C}}) & =Tr(z(1-qg_{E}(s))-(1-q))^{-1}{\bf{C}} \\
& \quad -(1-qg_{E}(s)){\bf{I_{N}}}) \\
& =N((1-qg_{E}(s))g_{E}(s)-1) \\
\end{split}
\end{equation} One also has $Tr(G_{E}(s){\bf{C}})=N(Z(s)g_{E}(s)-1)$, where $Z(s)=1-qg_{E}(s)$. So substitution of this and (\ref{29}) into (\ref{40}) yields:
\begin{equation}\label{41}
\begin{split}
\lim_{z\rightarrow \lambda _{i}- \eta }& Im[Tr(G_{C}(s){\bf{C}})]\\
&=N \pi \rho _{E}(\lambda_{i})((1-qh_{E}(\lambda_{i}))(\lambda_{i}-(1-q) \\
& \quad -2q\lambda_{i} h_{E}(\lambda_{i}))+q\varphi(\lambda_{i}))
\end{split}
\end{equation} Then by substituting function (\ref{41}) into (\ref{39}), we obtain the function (\ref{30}).
\section{Normal-inverse Gaussian Distribution}
The p.d.f of a NIG distribution is \cite{ou}:
\begin{equation}\label{42}
f(x)=\frac{ \alpha \delta K_{1}(\sqrt{\delta ^{2}+(x-\mu)^{2}})}{\pi \sqrt{\delta ^{2}+(x-\mu)^{2}}}e^{\delta \gamma +\beta (x-\mu)}
\end{equation} where $K_{1}(\cdot)$ denotes the modified Bessel function, and in this case $\alpha=1$, $\beta=1$, $\delta=1$, $\gamma=1$ and $\mu=-1$. The coefficients make the mean and variance of this NIG distribution become zero and $\sigma_{i}$.
\section{Supplementary Material}
This supplementary material explains what role the Kolmogolov limit plays in our work. Normally, the covariance matrix ${\bf{E=ZZ}}^{H} ={\bf{(H+G) (H+G)}}^{H}$ is estimated by $\mathbb{E}({\bf{E}})=\mathbb{E}({\bf{HH}}^{H} +{\bf{GH}}^{H} +{\bf{HG}}^{H} +{\bf{GG}}^{H})= {\bf{HH}}^{H}+{\bf{I_{N}}}$. However this well-known model is only accurate in an almost impossible condition: the number of samplings is much greater than the number of variables, i.e., $T>>N$. In the case of $N\sim O(T)$, this model is no longer fully trusted, since the eigenvalues distribution converges to the M-P law (\ref{36}) rather than to units \cite{WWW}. RMT exactly derives some deterministic properties of random matrices in the case of $N\sim O(T)$, by which we improve the WLS based state estimation.

\end{appendices}

 \ifCLASSOPTIONcaptionsoff
   \newpage
 \fi

\bibliographystyle{ieeetr} 
\bibliography{IEEEabrv,secebib}
\end{document}